\documentclass[12pt]{article}
\def\be{\begin{equation}}
\def\ee{\end{equation}}
\def\ie{{\it i.e.}}
\def\eg{{\it e.g.}}
\def\eq#1{(\ref{#1})}
\catcode`@=11
\@addtoreset{equation}{section}

\newdimen\z@ \z@=0pt
\def\m@th{\mathsurround=\z@}
\def\ialign{\everycr{}\tabskip\z@skip\halign} 
\def\eqalign#1{\null\,\vcenter{\openup\jot\m@th
  \ialign{\strut\hfil$\displaystyle{##}$&$\displaystyle{{}##}$\hfil
      \crcr#1\crcr}}\,}
\def\matrix#1{\null\,\vcenter{\normalbaselines\m@th
    \ialign{\hfil$##$\hfil&&\quad\hfil$##$\hfil\crcr
      \mathstrut\crcr\noalign{\kern-\baselineskip}
      #1\crcr\mathstrut\crcr\noalign{\kern-\baselineskip}}}\,}
\catcode`@=12
\font\black=msbm10 scaled\magstep1
\def\ld{\mathop{\cdots}\limits}
\def\mdot{\cdot}

\def\field #1{\hbox{{\black #1}}}

\def\R{{\hbox{{\field R}}}}

\def\g{{\cal G}}

\def\ind#1#2{{{#1_1}\ldots{#1_{#2}}}}
\def\ix#1#2#3{{{#1_{#2}}\ldots{#1_{#3}}}}
\renewcommand{\thefootnote}{\fnsymbol{footnote}}
\begin{document}
\rightline{To appear in Commun. Math. Phys.}
\rightline{hep-th/9605213}
\begin{center} 
\begin{large}
\vspace{0.5cm}
{\bf Higher-order simple Lie algebras}
\end{large}
\\[0.3cm]
J.A. de Azc\'{a}rraga
\footnote[1]{St. John's College Overseas Visiting Scholar.}
\footnotemark[2]
and 
J.C. P\'{e}rez Bueno 
\footnote[2]{On sabbatical (J.A.) leave and on leave of absence (J.C.P.B.) 
from Departamento de F\'{\i}sica Te\'orica and IFIC
(Centro Mixto Univ. de Valencia-CSIC) E--46100 Burjassot (Valencia), Spain.
\\
E-mails: j.azcarraga@damtp.cam.ac.uk (azcarrag@evalvx.ific.uv.es), 
pbueno@lie.ific.uv.es}
\\[0.3cm]
\begin{it}
Department of Applied Mathematics and Theoretical Physics, 
\\
Silver St., Cambridge, CB3-9EW, UK
\end{it}
\end{center}
\begin{abstract}
It is shown that the non-trivial cocycles on simple Lie algebras may be used 
to introduce antisymmetric multibrackets which lead to higher-order Lie 
algebras, the definition of which is given.
Their generalised Jacobi identities turn out to be 
satisfied by the antisymmetric tensors (or higher-order `structure 
constants') which characterise the Lie algebra cocycles.
This analysis allows us to present a classification of the higher-order simple 
Lie algebras as well as a constructive procedure for them.
Our results are synthesised by the introduction of a single, complete
BRST operator associated with each simple algebra.
\end{abstract}

\renewcommand{\thefootnote}{\arabic{footnote}}

\section{Introduction}

It is well known that, given 
$[X,Y]:=XY-YX$, the standard Jacobi identity (JI)
$[[X,Y],Z]+[[Y,Z],X]+[[Z,X],Y]=0$
is automatically satisfied if the product is associative (which will be assumed 
throughout).
For a Lie algebra $\g$, expressed by the Lie commutators 
$[X_i,X_j]=C_{ij}^k X_k$
in a certain basis 
$\{X_i\}\ i=1,\ldots,r={\rm dim}\g$,
the JI implies the Jacobi condition (JC)
\be
{1\over 2}\epsilon^{j_1j_2j_3}_{i_1i_2i_3}
C^\rho_{j_1 j_2}C^\sigma_{\rho j_3}=0\quad,
\label{jacid}
\ee
on the structure constants. 
Let $\g$ be simple. Using the Killing metric $k_{ij}=k(X_i,X_j)$
to lower and raise indices, the fully antisymmetric tensor
$C_{ijk}=C^s_{ij}k_{sk}=k([X_i,X_j],X_k)$
defines a non-trivial Lie algebra three-cocycle.
Since it is obtained from $k$, this three-cocycle is always present 
($H^3_0(\g,\R)\ne0$ for any $\g$ simple).
In fact, it is known since the classical work of Cartan, Pontrjagin, Hopf and 
others (see in particular 
\cite{CAR,PON,HOPF,HOD,CE,SAMEL,BORCHE,BOREL,BOTT}) 
that, from a topological point of view, the group manifolds of all simple 
compact groups are essentially equivalent to the products of odd spheres
\footnote{More precisely, if $G$ is a compact connected Lie group, $G$ has the 
({\it real}) cohomology or homology of a product of odd dimensional spheres.}, 
that $S^3$
is always present in these products and that the simple Lie algebra 
$\g$-cocycles 
are in one-to-one correspondence with bi-invariant de Rham cocycles 
on the associated compact group manifolds $G$.
The appearance of specific spheres $S^{2p+1}\ (p\ge 1)$ other than $S^3$ 
depends on the simple group considered.
This is due to the intimate relation between the order of the 
$l\!=$rank$\,\g$ 
primitive symmetric polynomials which can be defined on a 
simple Lie algebra, their $l$
associated generalised Casimir-Racah invariants 
\cite{RACAH,GEL,LCB,AK,GR,PP,NR,OKPA,SOK} 
\footnote{In the non-simple case the situation is more involved (see 
\cite{ABELLANAS}).}
and the topology of the 
associated simple groups, a fact which was found in the eighties to be a key 
in the understanding of non-abelian anomalies in gauge theories 
(see \cite{TJZW} for an 
account of the subject and \eg, \cite{AJ,KEP,OKPAII}).

By looking at the invariant symmetric polynomials on $\g$ we may obtain the 
higher-order cocycles of the Lie algebra cohomology. These cocycles will turn 
out to define $\g$-valued 
skew-symmetric brackets of even order $s$
satisfying a generalised Jacobi condition replacing \eq{jacid}.
Higher-order generalisations of Lie algebras, in the form of the {\it strongly 
homotopy Lie algebras} (SH) of Stasheff \cite{LAST,LAMA}, have recently 
appeared in physics.
This is the case of the (SH) algebra of products of closed string fields (see 
\cite{WZ,ZIE} and references therein), which involves multilinear, 
graded-commutative products of $n$ string fields satisfying certain `main 
identities' which also generalise the standard Jacobi identity.
The higher-order Lie algebras to be discussed in this paper satisfy, however, 
a generalised Jacobi condition which is a consequence of the assumed 
associativity of the product of algebra elements 
and which has also appeared in another, but 
related context \cite{APPB}.
As a result the definition of the skew-symmetric multibracket to be given 
in sec. 2 
permits, for each even $s$, the introduction of a coderivation 
$\partial_s$ of 
the exterior algebra $\wedge(\g)$ constructed on the Lie algebra $\g$.
In contrast, the `main identities' of the SH Lie algebras \cite{LAST,LAMA}
(some detailed expressions can be found in \cite{EJ})
involve a further extension of our generalised Jacobi identities which in 
effect describes how the products fail to satisfy them
and how the various $\partial_s$ involved in the main identities 
fail separately to be a coderivation.
Our extended higher-order algebras are thus a particular case of the SH Lie 
algebras in which only one of the $\partial_s$ is non-zero
\footnote{
Note added: This is also the case of the `$k$-algebras' \cite{HW}.
We thank P. Hanlon for sending us this reference.}.
We shall now show how introduce them (sec. 2) and present their Cartan--like 
classification in the simple case (secs. 3,4).
In sec. 5 we shall describe our results by introducing
the complete BRST operator associated with a simple Lie 
algebra; some comments concerning applications and extensions will be made in 
sec. 6.

\section{Multibrackets and higher-order Lie algebras}

Higher-order Lie algebras may be defined by introducing a 
suitable generalisation of the Lie bracket by means of

\medskip
\noindent
{\bf Definition 2.1}\quad ({\it $s$-bracket})

Let $s$ be even.
A $s$-{\it bracket} or skew-symmetric Lie multibracket is a Lie algebra 
valued $s$-linear skew-symmetric mapping 
$\g\times\ld^s\times\g\to \g$,
\be
(X_{i_1},X_{i_2},\ldots,X_{i_s})\mapsto
[X_{i_1},X_{i_2},\ldots,X_{i_s}]=
{\omega_{\ind is}}^\sigma_\mdot X_\sigma\quad,
\label{defpars}
\ee
where the constants ${\omega_{\ind is}}^\sigma_\mdot$  satisfy the condition
\be
\epsilon^\ind j{2s-1}_\ind i{2s-1}
{\omega_{\ind j s}}^\rho_\mdot
{\omega_{\rho \ix j{s+1}{2s-1} }}^\sigma_\mdot=0\quad.
\label{jaccond}
\ee
The ${\omega_{\ind is}}^\sigma_\mdot$ will be called {\it higher-order 
structure constants}, and condition \eq{jaccond} will be referred to as the 
{\it generalised Jacobi condition} (GJC); for $s=2$ it gives the ordinary JC 
\eq{jacid}.

\medskip
\noindent
{\it Remark.}\quad
Although we shall only consider here the case of Lie algebras,
this definition (as others below) is more general.

\medskip
The GJC \eq{jaccond} is clearly a consistency condition for 
\eq{defpars}. From now on $X_i$ will denote both the algebra basis elements 
and its representatives in a faithful representation of $\g$.
Let now 
$[X_{i_1},X_{i_2},\allowbreak\ldots,X_{i_n}]$, $n$ arbitrary,
be defined  by
\be
[X_{i_1},X_{i_2},\ldots,X_{i_n}]=
\sum_{\sigma\in S_n}(-1)^{\pi(\sigma)}
X_{i_{\sigma(1)}}X_{i_{\sigma(2)}}\ldots X_{i_{\sigma(n)}}
\quad,
\label{defmulti}
\ee
where $\pi(\sigma)$ is the parity of the permutation $\sigma$ and the 
(associative) products on the $r.h.s.$ are well defined as products of 
matrices or as elements of ${\cal U}(\g)$.
Then, the following Lemma holds:

\medskip
\noindent
{\bf Lemma 2.1}\quad

Let $[X_1,\ldots, X_n]$, be as in \eq{defmulti} above. 
Then, for $n$ even,
\be
{1\over (n-1)!}{1\over n!}\sum_{\sigma\in S_{2n-1}} (-1)^{\pi(\sigma)}
[[X_{\sigma(1)},\ldots,X_{\sigma(n)}],X_{\sigma(n+1)},\ldots,X_{\sigma(2n-1)}] 
=0
\label{genjacid}
\ee
is an identity, the {\it generalised Jacobi identity} (GJI) which for 
\eq{defpars} 
implies the GJC \eq{jaccond}; for $n$ odd, the 
$l.h.s.$ is proportional to $[X_1,\ldots,X_{2n-1}]$.

\medskip
\noindent
{\it Proof}:

Let $Q_p$ be the antisymmetriser for the symmetric group 
$S_p$ (\ie, the pri\-mi\-ti\-ve `idempotent' 
$[Q_p^2=p!Q_p]$ in the Frobenius algebra of $S_p$, associated with the fully 
antisymmetric Young tableau).
The sum in \eq{genjacid} contains $C^{n-1}_{2n-1}$ different terms
($(2n-1)!/(n-1)!n!=C^{n-1}_{2n-1}$).
Consider the first of these,
$[[X_1,\ldots,X_n],X_{n+1},\ldots,\allowbreak X_{2n-1}]$.
Its full expansion contains $(n!)^2$ terms, which may be written as the sum 
of $n$ terms
\be
\eqalign{
[[X_1,\ldots,X_n]&
,X_{n+1},\ldots,X_{2n-1}]=
Q_n(X_1X_2\ldots X_n)Q_{n-1}(X_{n+1}X_{n+2}\ldots X_{2n-1})
\cr
&
-Q_{n-1}(X_{n+1} Q_n(X_1X_2\ldots X_n) X_{n+2}\ldots X_{2n-1})
\cr
&
+Q_{n-1}(X_{n+1} X_{n+2} Q_n(X_1X_2\ldots X_n)X_{n+3}\ldots X_{2n-1})
\cr
&
+\ldots+(-1)^{n-2}Q_{n-1}(X_{n+1} X_{n+2} \ldots X_{2n-2} Q_n(X_1X_2\ldots X_n) 
X_{2n-1})
\cr
&
+ (-1)^{n-1} Q_{n-1}(X_{n+1} X_{n+2}\ldots X_{2n-1} )
Q_n(X_1X_2\ldots X_n)\quad,
\cr}
\label{proofjacobi}
\ee
where the antisymmetriser $Q_n$ $[Q_{n-1}]$ acts on the $n$ $[n-1]$ indices 
$(1,\ldots,n)$ $[(n+1,\ldots,2n-1)]$ {\it only}.
This sum may be rewritten as
\be
\eqalign{
&
Q_nQ_{n-1}\{e+(-1)^n(1,n+1)+(1,n+1)(2,n+2)+
(-1)^n(1,n+1)(2,n+2)\mdot
\cr
&\mdot (3,n+3)+\ldots+(-1)^n(1,n+1)(2,n+2)\ldots(n-2,2n-2)+
\cr
&
\phantom{\times}(1,n+1)\ldots(n-1,2n-1)\}X_1\ldots X_{2n-1}
\quad,
\cr}
\label{moreproofjacobi}
\ee
where $(i,j)$ indicates the transposition in $S_{2n-1}$ which interchanges the 
indices $i,j$;
thus, all the signs in \eq{moreproofjacobi} are positive for $n$ even, and they alternate for $n$ odd 
according to the parity of the accompanying permutation.

Numerical factors apart, the $l.h.s$ of \eq{genjacid} is the result of the 
action of the 
$S_{2n-1}$ antisymmetriser in $(2n-1)$ indices, $Q_{2n-1}$, on 
\eq{proofjacobi} or 
\eq{moreproofjacobi}.
Since
$\sigma Q_{2n-1}=(-1)^{\pi(\sigma)}Q_{2n-1}\, \forall\sigma\in S_{2n-1}\,,$ it 
turns out that 
$Q_{2n-1}(Q_n Q_{n-1})\propto Q_{2n-1}$.
Thus, only the action of $Q_{2n-1}$ on the curly bracket in 
\eq{moreproofjacobi} 
has to be considered.
Since its permutations are half even and half odd, it becomes identically zero 
for $n$ even and proportional to $Q_{2n-1}$ for $n$ odd, {\it q.e.d.}

\medskip
Lemma 2.1 shows that the higher-order bracket may be defined, 
as the Lie bracket, by the skew-symmetric product of an (even) number of 
generators.
By analogy with the standard  Lie algebra ($s=2$) case, we may now
give the following

\medskip
\noindent
{\bf Definition 2.2}\quad ({\it Higher-order Lie algebra})

Let $\g$ be a Lie algebra.
A higher-order Lie algebra on $\g$ is the algebra defined by the 
$s$-bracket \eq{defpars},
where the higher-order structure constants satisfy the generalised Jacobi 
condition \eq{jaccond}.

\medskip
Multibrackets appear naturally if we use for the basis $X_i$  of $\g$ a set 
of left-invariant vector fields (LIVF) on the group manifold\footnote{
On $G$, a vector field $X_i$ is expressed as 
$X_i^j(g)\partial/\partial g^j\,,\,j=1,\ldots,r\;,$ 
where $g^i$ are local coordinates of $G$ at the unity.}
of the Lie group $G$ associated with $\g$. 
Then, the exterior algebra $\wedge(G)$ may be identified as the exterior 
algebra of the LI contravariant, skew-symmetric tensor fields on $G$
obtained by taking the exterior products of LIVF's with constant 
coefficients; this is analogous to the exterior algebra of LI covariant 
tensor fields (LI forms) on $G$.
Then, in analogy
with the exterior derivative of a LI $q$-form 
$\omega\in\wedge_q(G)$,
an exterior {\it coderivation}
$\partial:\wedge^q(G)\to\wedge^{q-1}(G)\,,\,\partial^2=0,$
may be introduced by taking 
\be
\partial(X_1\wedge\ldots\wedge X_q)=
\sum^q_{\scriptstyle l=1 \atop {\scriptstyle l<k}}(-1)^{l+k+1}[X_l,X_k]\wedge 
X_1\wedge\ldots\widehat X_l\ldots\widehat X_k\ldots\wedge X_q\quad.
\label{coder}
\ee
For instance, on 
$X_{i_1}\wedge X_{i_2}\wedge X_{i_3}\in \wedge^3(G)$, the statement
$\partial^2(X_{i_1}\wedge X_{i_2}\wedge X_{i_3})=0$ 
is nothing but the standard Jacobi identity.

If we now define
$\partial_2(X_{i_1}\wedge X_{i_2})=
\epsilon^{j_1 j_2}_{i_1 i_2} 
X_{j_1}X_{j_2}=[X_{i_1},X_{i_2}],$ 
the coderivation $\partial$ above corresponds to
$\partial_2:\wedge^q(G)\to\wedge^{q-1}(G)$.
This may now be extended to a general {\it even} coderivation 
$\partial_s$,
$\partial_s:\wedge^q(G)\to\wedge^{q-(s-1)}(G)\,,\, \partial_s^2=0\,$:

\medskip
\noindent
{\bf Definition 2.3}\quad ({\it coderivation $\partial_s$})

Let $s$ be even.
The mapping $\partial_s:\wedge^s(G)\to\wedge^1(G)\sim\g$ given by 
$\partial_s:X_1\wedge\ldots\wedge X_s\mapsto [X_1,\ldots,X_s]$, where the 
$s$-bracket is given by Def. 2.1, may 
be extended to a higher-order coderivation 
$\partial_s:\wedge^n(G)\to\wedge^{n-s+1}$ 
by
\be
\partial_s(X_1\wedge\ldots\wedge X_n)=
{1\over s!(n-s)!}\epsilon^{\ind i n}_{1\, \ldots\, n}
\partial_s(X_{i_1}\wedge\ldots\wedge X_{i_s})\wedge X_{i_{s+1}}\wedge\ldots 
\wedge X_{i_n}\;,
\label{extended}
\ee
with $\partial_s:\wedge^n(G)=0$ for $s>n$.
It follows from \eq{jaccond} that $\partial_s^2=0$. 

\medskip
For $s=2$, eq. \eq{extended} reduces to \eq{coder}.
On $X_{i_1}\wedge\ldots\wedge X_{i_7}\in\wedge^7(G)$, for instance,
$\partial_4^2=0$ leads to the GJI which must be satisfied by a 4-$th$ order 
Lie algebra.
As mentioned, these higher-order algebras are particular cases of the  
strongly homotopy algebras \cite{LAST,LAMA} of recent relevance in string 
field theory (see \cite{ZIE}).
We shall now give explicit examples of higher-order algebras and, as a 
result, provide the classification of all higher-order simple Lie algebras.


\section{Higher-order simple Lie algebras. The case of $su(n)$}

Let $\g$ be now a simple Lie algebra.
In what follows, we shall also assume $\g$ to be compact 
(although compactness is not essential in many reasonings below) 
so that the non-degenerate Killing matrix $k_{ij}$ may 
be taken as the unity $\delta_{ij}$
after suitable normalization of the generators. 

As mentioned, there are $l$ primitive invariant
polynomials for each simple algebra of rank $l$ which are in turn related
to the Casimir-Racah operators of the algebra
\cite{WEYL,RACAH,LCB,AK,GR,NR,OKPA,SOK,EK,LJB}, to the Lie algebra 
cohomology for the trivial action and to the topology
and de Rham cohomology of the associated simple compact Lie
group \cite{CAR,HOPF,HOD,CE,SAMEL,BORCHE,BOREL}.
We now use this fact to provide a classification of the possible
higher-order simple Lie algebras.
Given a simple Lie algebra $\g$, the orders $m_i$ of the $l$ invariant
polynomials (or of the generalised Casimir invariants) and of the $l$ cocycles
(or bi-invariant forms on the corresponding compact group $G$) 
are given by the following table
 
\begin{center}
\medskip
\noindent
\vbox{\tabskip=0pt\offinterlineskip
\def\tablerule{\noalign{\hrule}}
\halign to 15.9cm{\strut#&\vrule#\tabskip=0.2em plus 0.2em&
\hss $#$ \hss &
\vrule # &
\hss $#$ \hss &
\vrule # &
\hss $#$ \hss &
\vrule # &
\hss $#$ \hss &
\vrule #\tabskip=0pt\cr\tablerule
&&\g &&  \hbox{algebra\ dimension}&& \hbox{order\ of\ invariants} && 
\hbox{order of $\g$-cocycles}
& \cr
&& && r=\hbox{dim}\g && m_1,\ldots,m_l && (2m_1-1),\ldots,(2m_l-1) 
&\cr\tablerule
&& A_l && (l+1)^2-1\ [l\ge 1]&& 2,3,\ldots,l+1 && 3,5,\ldots, 2l+1
&\cr\tablerule
&& B_l && l(2l+1)\ [l\ge 2] && 2,4,\ldots,2l && 3,7,\ldots, 4l-1
&\cr\tablerule
&& C_l && l(2l+1)\ [l\ge 3] && 2,4,\ldots,2l && 3,7,\ldots, 4l-1
&\cr\tablerule
&& D_l && l(2l-1)\ [l\ge 4] && 2,4,\ldots,2l-2,\,l && 3,7,\ldots,4l-5,\,2l-1
&\cr\tablerule
&& G_2 && 14 && 2,6 && 3,11
&\cr\tablerule
&& F_4 && 52 && 2,6,8,12 && 3,11,15,23
&\cr\tablerule
&& E_6 && 78 && 2,5,6,8,9,12 && 3,9,11,15,17,23
&\cr\tablerule
&& E_7 && 133 && 2,6,8,10,12,14,18 && 3,11,15,19,23,27,35
&\cr\tablerule
&& E_8 && 248 && 2,8,12,14,18,20,24,30 && 3,15,23,27,35,39,47,59
&\cr\tablerule
}}
\end{center}

\noindent
{\it 
Dimension of the Casimir-Racah invariants and Lie algebra cocycles 
for $\g$ simple}.

\medskip
\noindent
We see that $\sum_{i=1}^l(2m_i-1)=r$. 

\medskip
\noindent
{\bf Definition 3.1}\quad ({\it Higher-order simple Lie algebras})

A higher-order simple Lie algebra associated with a simple Lie algebra $\g$ is 
the higher-order algebra defined by a primitive $\g$-cocycle (of order $>3$) 
on $\g$.

\medskip
Thus, to find the higher-order simple Lie algebras one has to look for
the invariant polynomials on them.
For the compact forms on these groups, the cocycle orders are also the 
dimensions of the primitive de Rham cycles (odd spheres) to which the group 
manifolds are essentially equivalent.
We shall now find explicit realizations of these algebras.

Consider first the case of 
$su(n)\,,\,n\ge 3$ with $k_{ij}\sim\delta_{ij}$
(there are no higher-order simple Lie algebras on $su(2)$).
In terms of its structure constants (for hermitian generators $T_i$) 
$[T_i,T_j]=iC_{ijk}T_k$, the anticommutator of two $n\times n\; su(l+1)$
matrices may be expressed as
$\{T_i,T_j\}=c\delta_{ij} + d_{ijk}T_k$ (with $c=1/n$,
Tr$(T_iT_j)={1\over 2}\delta_{ij}$).
The $d_{ijk}\propto {\rm Tr}(T_i\{T_j,T_k\})$ term (absent for $su(2)$) 
is the first example of a symmetric invariant polynomial 
(of $3rd$ order\footnote{
For the properties of the $d$-tensors see \cite{SUDBERY}.}) beyond the
Killing tensor $k_{ij}$ (see Table).
Invariant, symmetric polynomials are given by the symmetric traces 
(sTr)
of products of $su(n)$ generators (cf. the theory of characteristic classes).
Let us then consider the next case, $m_3=4$.
The coordinates of this fourth-order polynomial $k_{i_1 i_2 i_3 i_4}$ are
given by 
sTr$(T_{i_1} T_{i_2} T_{i_3} T_{i_4})$ or (ignoring numerical
factors) by
Tr($s(T_{i_1} T_{i_2} T_{i_3})
T_{i_4})\propto$Tr($s(\{\{T_{i_1},T_{i_2}\},T_{i_3}\})T_{i_4})\propto
(d_{(i_1 i_2 l}d_{l i_3) i_4}+ 2c\delta_{(i_1 i_2}\delta_{i_3)i_4})$
where $s$ symmetrises the $i_1,i_2,i_3$ indices.
Thus, we may take
\be
\eqalign{
k_{i_1 i_2 i_3 i_4}=
&
d_{i_1 i_2 l}d_{l i_3 i_4}+d_{i_1 i_3 l}d_{l i_2 i_4}+
d_{i_1 i_4 l}d_{l i_2 i_3}
\cr
&
+2c(\delta_{i_1 i_2}\delta_{i_3 i_4}+\delta_{i_1 i_3}\delta_{i_2 i_4}+
\delta_{i_1 i_4}\delta_{i_2 i_3})\quad.
\cr}
\label{fourthorder}
\ee
Clearly, the last term will not generate a primitive 4-$th$ order 
Casimir operator\footnote{Notice that $l\ge 3$ for $k_{i_1 i_2 i_3 i_4}$ to be 
primitive. For $su(3)$ the identity 
$d_{(i_1 i_2 l} d_{i_3 ) i_4 l} = {1\over 3} 
\delta_{(i_1 i_2} \delta_{i_3) i_4}$, where the brackets mean symmetrisation, 
precludes eq. (\ref{fourthorder}) from producing a primitive fourth-order 
invariant.
Similar type relations hold for higher ranks \cite{SUDBERY} (see also 
\cite{BAIS}, 
where higher order Casimir operators were used to introduce the so-called 
Casimir ${\cal W}$-algebras).}, since it
is proportional to the square of the second order one, $(I_2)^2$.
Eq. \eq{fourthorder} reflects the well known ambiguity in the selection of
the higher-order Casimirs for the simple Lie algebras  (see, \eg,
\cite{LCB,GR,SOK,BW}). 
The first part, which generalises easily up to
$k_\ind in$ leads to the form of the Casimir-Racah operator $I_n$ given in 
\cite{AK}.

We are now in a position to introduce all $A_l$ higher-order simple 
Lie algebras

\medskip
\noindent
{\bf Theorem 3.1}\quad ({\it Higher-order $A_l$ Lie algebras})

Let $X_i$ a basis of $A_l$, $i=1,\ldots, (l+1)^2-1$.
Then, the even multibracket
\be
[X_{i_1},\ldots,X_{i_{2m-2}}]:=
\epsilon^\ind j{2m-2}_\ind i{2m-2}X_{j_1}\ldots X_{j_{2m-2}}
\label{evenmultibracket}
\ee
is $\g$-valued and defines a higher-order simple Lie algebra
\be
[X_{i_1},\ldots,X_{i_{2m-2}}]={\omega_\ind i{2m-2}}^\sigma_\mdot X_\sigma
\quad,
\label{cocycle}
\ee
where the higher-order structure constants 
${\omega_\ind i{2m-2}}^\sigma_\mdot$ associated
to the invariant polynomial $k_\ind i m$ are given by the skew-symmetric tensor
\be
\omega_{\ind i{2m-2}\sigma}=\epsilon^\ix j2{2m-2}_\ix i2{2m-2}
C^{l_1}_{i_1 j_2}\ldots C^{l_{m-1}}_{j_{2m-3}j_{2m-2}}k_{\ind l{m-1}\sigma}
\quad,
\label{defcocycle}
\ee
which defines a non-trivial $(2m-1)$-cocycle for $su(l+1)\,,\,3\le m\le l+1$
($m=2$ is the standard Lie algebra).

Before presenting a general proof, let us illustrate the theorem in the two
simplest cases.
For $m=2$ eq. \eq{defcocycle} reads
\be
\omega_{i_1 i_2 \sigma}=\delta^{j_2}_{i_2}C^{l_1}_{i_1j_2}k_{l_1\sigma}=
k([X_{i_1},X_{i_2}],X_\sigma)
\quad,
\label{secondorder}
\ee
and the $\omega_{i_1 i_2 \sigma}$ are the standard structure
constants of $\g$.
Thus, the $m=2$ (lowest) polynomial corresponds to the ordinary ($su(n)$, in
this case) Lie algebra commutators.

Let $m=3$.
If $d$ denotes the symmetric polynomial, eq. \eq{defcocycle} gives
\be
\eqalign{
\omega_{i_1 i_2 i_3 i_4\sigma}=
&
\epsilon^{j_2 j_3 j_4}_{i_2 i_3 i_4} C^{l_1}_{i_1 j_2}C^{l_2}_{j_3 j_4}
d_{l_1 l_2 \sigma}=
\cr
=&
\epsilon^{j_2 j_3 j_4}_{i_2 i_3 i_4} 
d([X_{i_1},X_{j_2}],[X_{j_3},X_{j_4}],X_\sigma)
\quad,\cr}
\label{examplei}
\ee
which is the expression of the fully antisymmetric five-cocycle. On the
other hand,
\be
\eqalign{
[X_{i_1},X_{i_2},X_{i_3},X_{i_4}]
& =
\epsilon^{j_1 j_2 j_3 j_4}_{i_1 i_2 i_3 i_4}X_{j_1}\ldots X_{j_4}=
{1\over 2^2}\epsilon^{j_1 j_2 j_3 j_4}_{i_1 i_2 i_3 i_4}
[X_{j_1},X_{j_2}][X_{j_3},X_{j_4}]
\cr
&
={1\over 2^2}\epsilon^{j_1 j_2 j_3 j_4}_{i_1 i_2 i_3 i_4}
C^{k}_{j_1 j_2}C^{l}_{j_3 j_4}X_k X_l
\quad.\cr}
\label{exampleii}
\ee
Taking into account that
$\epsilon^{j_1 j_2 j_3 j_4}_{i_1 i_2 i_3 i_4}
C^{k}_{j_1 j_2}C^{l}_{j_3 j_4}$ is symmetric in $k,l$ this is equal to 
\be
{1\over 2^3}
\epsilon^{j_1 j_2 j_3 j_4}_{i_1 i_2 i_3 i_4}
C^{l_1}_{j_1 j_2}C^{l_2}_{j_3 j_4}
(d_{l_1l_2\sigma} X_\sigma + c\delta_{l_1l_2})\quad.
\label{exampleiii}
\ee
The term in $c$ may be dropped since, for each $j_4$, it is proportional
to the antisymmetrised sum 
$C_{j_1 j_2 l}C^l_{j_3 j_4}$ in $j_1,j_2,j_3$ which is zero by the
Jacobi identity.
Using now that
\be
\epsilon^{j_1 j_2 j_3 j_4}_{i_1 i_2 i_3 i_4}=\sum^4_{s=1}(-1)^{s+1}
\delta_{i_1}^{j_{s}}\epsilon^{j_1 \ldots\widehat j_s\ldots j_4}_{i_2 i_3 i_4}
\quad,
\label{epsilonprop}
\ee
it is easy to see that all the terms in \eq{epsilonprop} give the same
contribution for the remaining $d$ term in \eq{exampleiii}.
Hence, the fourth-commutator
\be
[X_{i_1},X_{i_2},X_{i_3},X_{i_4}]=
{1\over 2}
\epsilon^{j_2 j_3 j_4}_{i_2 i_3 i_4} C^{l_1}_{i_1 j_2}C^{l_2}_{j_3 j_4}
d_{l_1 l_2 \sigma}X_\sigma
\label{exampleiv}
\ee
is indeed of the form \eq{examplei}, and it may be checked explicitly that it
is in $su(3)$. 

The proof of Theorem 3.1 requires now the following simple

\medskip
\noindent
{\bf Lemma 3.1}\quad

If $k_\ind lm$ is an ad-invariant, symmetric polynomial on a simple Lie
algebra $\g$,
\be
\epsilon^\ind j{2m}_\ind i{2m} 
C^{l_1}_{j_1 j_2}\ldots C^{l_m}_{j_{2m-1} j_{2m}}
k_\ind lm=0
\quad.
\label{simplelemma}
\ee

\medskip
\noindent
{\it Proof}:

First, we note that the ad-invariance 
condition of the $m$-tensor $k$ may be expressed in coordinates by
\be
\sum_{s=1}^m C^{k}_{j_{2m-1} l_s} 
k_{l_1\ldots l_{s-1} k l_{s+1}\ldots l_m}=0\quad.
\label{invariance}
\ee
Hence, replacing $C^{l_m}_{j_{2m-1} j_{2m}}k_\ind lm$ in the $l.h.s.$ of
\eq{simplelemma}\ by the other terms in \eq{invariance}\ we get
\be
\epsilon^\ind j{2m}_\ind i{2m} 
C^{l_1}_{j_1 j_2}\ldots C^{l_{m-1}}_{j_{2m-3} j_{2m-2}}
(\sum_{s=1}^{m-1}C^{k}_{j_{2m-1} l_s} k_{l_1\ldots l_{s-1} k l_{s+1}\ldots 
l_{m-1}\,j_{2m}})
\quad,
\ee
which vanishes since all terms in the sum include products of the form
$C^s_{jj'}C^k_{sj''}$ antisymmetrised in $j,j',j''$, which are zero due to
the standard JC \eq{jacid}, {\it q.e.d.}

\medskip
To prove now Theorem 3.1, we write the $(2m-2)$ bracket as
\be
\eqalign{
[X_{i_1},\ldots, X_{i_{2m-2}}]=
&
{1\over 2^{m-1}}\epsilon^\ind j{2m-2}_\ind i{2m-2} 
[X_{j_1},X_{j_2}]\ldots[X_{j_{2m-3}},X_{j_{2m-2}}]=
\cr
&
{1\over 2^{m-1}}\epsilon^\ind j{2m-2}_\ind i{2m-2} 
C^{l_1}_{j_1 j_2}\ldots C^{l_{m-1}}_{j_{2m-3} j_{2m-2}}
{1\over (m-1)!}s(X_{l_1}\ldots X_{l_{m-1}})
\cr}
\label{prooftheoremi}
\ee
where we have used (cf. the $m=3$ case) that
$\epsilon^\ind j{2m-2}_\ind i{2m-2}
C^{l_1}_{j_1 j_2}\ldots C^{l_{m-1}}_{j_{2m-3} j_{2m-2}}$
is symmetric in $l_1,\ldots,l_{m-1}$ to introduce the symmetrised product
of generators, which in turn may be replaced, adding the appropriate
factors, by
$s(\{\{\ldots\{X_{l_1},X_{l_2}\},X_{l_3}\},\allowbreak\ldots,X_{l_{m-1}}\})$.
Using that $\{X_i,X_j\}=c\delta_{ij}+d_{ijk}X_k$ 
in the expression of the nested
anticommutators, we then conclude that it has the form
\be{\rm (factors)}s(X_{l_1}\ldots X_{l_{m-1}})=
{\tilde k}_{l_1\ldots l_{m-1}\mdot}^{\phantom{l_1\ldots l_{m-1}}\sigma} 
X_\sigma+\hat k_{l_1\ldots l_{m-1}}1\quad.
\label{symproduct}
\ee
By Lemma 3.1, the second term does not contribute to \eq{prooftheoremi}\ 
because 
$\hat k$ is an invariant polynomial of $(m-1)$-order.
On the other hand since
Tr$(s(X_{l_1}\ldots X_{l_{m-1}})X_\sigma)\propto\,
$sTr$(X_{l_1}\ldots X_{l_{m-1}}X_\sigma)$,
we conclude that 
$\tilde k_{l_1\ldots l_{m-1}\sigma}$ is an invariant symmetric $m$-th
order polynomial.
Absorbing all numerical factors in $\tilde k$ and renaming it as $k$, we
find that the $(2m-2)$-commutator in \eq{prooftheoremi}\ is given by
\be
{1\over (2m-2)}\epsilon^\ind j{2m-2}_\ind i{2m-2} 
C^{l_1}_{j_1 j_2}\ldots C^{l_{m-1}}_{j_{2m-3} j_{2m-2}}
{k_{l_1\ldots l_{m-1}}}^\sigma_\mdot X_\sigma=
{\omega_\ind i{2m-2}}^\sigma_\mdot X_\sigma
\ee
\ie, by the $(2m-1)$-cocycle \eq{defcocycle}. 
Since \eq{evenmultibracket} 
is given by the product of associative operators, 
the GJC \eq{jaccond} follows from the GJI \eq{genjacid}, {\it q.e.d.}
Equivalently, one may show that the cocycle condition for 
$\omega_{\ind i{2m-2}\sigma}$ guarantees that the GJC is satisfied (see the 
Remark after Th. 5.1 below).
This establishes the connection between Lie algebra cohomology cocycles and 
higher-order Lie algebras.

\section{Higher-order orthogonal and symplectic algebras}

We now extend to the 
$B_l\ (l\ge 2),\ C_l\ (l\ge 3),\ D_l\ (l\ge 4)$ 
series the considerations in sec. 3 for
$A_l$.
First we notice that for all of them the third-order symmetric polynomial
is absent and that only for the even orthogonal algebra $D_l$
(and odd $l$) we may have an
odd-order invariant polynomial.
We shall ignore this case for a moment, 
and look first for the even-order symmetric polynomials.
Let us realise the generators of the above algebras
in terms of the $n\times n$
matrices of the defining representation, 
where $n=(2l+1,2l,2l)$ for $(B_l,C_l,D_l)$ respectively.
These matrices $T$ have all in common the metric preserving defining property
$Tg=-gT^t$, where $g$ is the $n\times n$ unit matrix for the orthogonal 
algebras and the symplectic metric for $C_l$.
If we define the symmetric third-order anticommutator by
\be
\{T_1,T_2,T_3\}=\sum_{\sigma\in S_3}T_{\sigma(1)}T_{\sigma(2)}T_{\sigma(3)}
\equiv s(T_1T_2T_3)\quad,
\label{threeanticom}
\ee 
it is trivial to check that 
$\{T_1,T_2,T_3\}g=-g\{T_1,T_2,T_3\}^t$
so that
$\{T_1,T_2,T_3\}\in\g\ (\g=so(2l+1)\,,\,sp(2l)\,{\rm or}\,so(2l))$.
Notice that such a relation cannot be satisfied for the ordinary
anticommutator, and that in general requires {\it odd}-order anticommutators 
in order to preserve the minus sign in the $r.h.s.$
Note also the absence of the identity matrix in the $r.h.s.$ of the
odd-order anticommutator, which was allowed for $A_l$.
Let then
$\{T_{i_1},T_{i_2},T_{i_3}\}={k_{i_1i_2i_3}}^\sigma_\mdot T_\sigma$.
Extending this result to the arbitrary odd case, we find

\medskip
\noindent
{\bf Lemma 4.1}

The symmetrised product of an odd number of $n\times n$ matrix generators of 
$so(2l+1)\,,\,sp(2l)$ or $so(2l)$ is also an element of these algebras
which is determined by the associated invariant symmetric polynomial.

\medskip
\noindent
{\it Proof}:
\be
\eqalign{
s(T_{i_1}T_{i_2}T_{i_3}T_{i_4}\ldots T_{i_{2p-1}})=
&
{1\over 6^{p-1}}s(\{\ldots\{\{T_{i_1},T_{i_2},T_{i_3}\}T_{i_4},T_{i_5}\},
\ldots,T_{i_{2p-2}},T_{i_{2p-1}}\})
\cr
=
&{1\over 6^{p-1}}
s({k_{i_1i_2i_3}}^{\alpha_1}_\mdot{k_{\alpha_1 i_4i_5}}^{\alpha_2}_\mdot\ldots
{k_{\alpha_{p-2} i_{2p-2}i_{2p-1}}}^\beta_\mdot T_\beta)\quad.
\cr}
\label{orthogonal}
\ee
Since $s$ symmetrises all $i_1,i_2,\ldots,i_{2p-1}$ indices we may write
this as
\be
\{T_{i_1},\ldots,T_{i_{2p-1}}\}={k_{i_1\ldots i_{2p-1}}}^\sigma_\mdot T_\sigma
\quad,
\label{invpolynomial}
\ee
and identify $k$ with the invariant symmetric polynomial of even
$2p$ (see Table) since 
Tr($\{T_{i_1},\ldots\allowbreak,T_{i_{2p-1}}\}T_\sigma)$ is equal to
\be
{\rm sTr}(T_{i_1}\ldots T_{i_{2p-1}}T_\sigma)=
k_{i_i\ldots i_{2p-1}\sigma}
\quad,
\label{invpol}
\ee
{\it q.e.d.}

This now leads to the following

\medskip
\noindent
{\bf Theorem 4.1}

Let $\g$ be a simple orthogonal or symplectic algebra.
Let $k_\ind i{2p}$ be as in \eq{invpol} for $2\le p\le l$ ($B_l,C_l$) and
$2\le p\le l-1$ ($D_l$).
Then, the even $(4p-2)$ bracket defined as in \eq{evenmultibracket} 
defines a
higher-order orthogonal or symplectic algebra, the structure constants of
which are given by the Lie algebra $(4p-1)$-cocycles associated with the 
symmetric invariant polynomials on $\g$.

\medskip
\noindent
{\it Proof}:

It suffices to use Lemma 4.1 and to insert \eq{invpolynomial} in expression 
\eq{prooftheoremi}.
As a result, the $(4p-1)$-cocycle is given again by \eq{defcocycle} where the
$k_{i_i\ldots i_{2p-1}\sigma}$ 
is now found in \eq{invpol}, {\it q.e.d.}

\medskip
Let us consider now the order $l$ invariant for $so(2l)$.
The reasonings before Lemma 4.1 show that, for $l$ odd, the order $l$ 
invariant polynomial cannot be obtained from the symmetric trace of $(l-1)$ 
$2l\times 2l\ T$'s, since the symmetrised bracket of an even number of $T$'s 
cannot be expressed as a linear combination of the 
$2l\times 2l$ matrix generators of $so(2l)$.
It is well known, however, that for $so(2l)$ there is an order $l$ (even or 
odd) invariant polynomial (which gives the Euler class of a real oriented 
vector bundle with even-dimensional fibre) 
which comes from the Pfaffian, since $Pf(ATA^{t})=Pf(T)$ for 
$A\in SO(2l)$.
Using pairs of indices to relabel the generators 
$T_i\ i=1,\ldots,({2l \atop 2})$ 
as $T_{\mu \nu}=-T_{\nu\mu}\ ,\ \mu,\nu=1,\ldots,2l$, the order $l$ invariant 
(corresponding to the last one in the Table for $D_l$) is given by
\be
Pf(T)={(-1)^l\over 2^l l!}
\epsilon^{\mu_1\nu_1\ldots \mu_l\nu_l}_{1\;\ldots\;2l}
T_{\mu_1\nu_1} T_{\mu_2\nu_2}\ldots T_{\mu_l\nu_l}\quad.
\label{pfaff}
\ee
The antisymmetric tensor   
$\epsilon_{\mu_1\nu_1\mu_2\nu_2\ldots \mu_l\nu_l}$ defining the invariant is 
symmetric under the exchange of
{\it pairs} of indices $(\mu_i\nu_i)\ i=1,\ldots,l$.
Although it cannot be obtained as the symmetric trace of a product of 
$2l\times 2l$ generators it may be obtained again in the standard way if we 
use an appropriate spinorial representation for $so(2l)$.
This means  that the previous arguments may be also carried through to the 
$l$-$th$ order invariant of the $D_l$ algebra.
To see it explicitly, consider the $2^l$-dimensional Clifford algebra
$\{\Gamma_\mu,\Gamma_\nu\}=2\delta_{\mu\nu}\ (\mu,\nu=1,\ldots,2l)$.
The $({2l\atop 2})$ $Spin(2l)$ generators are given by 
$\Sigma_{\mu \nu}={i\over 2}[\Gamma_\mu,\Gamma_\nu]$, and the $\Gamma_{2l+1}$ 
matrix by
$\Gamma_{2l+1}={i^l\over (2l)!}\epsilon_{\ind \mu{2l}}\Gamma_{\mu_1}\ldots
\Gamma_{\mu_{2l}}$; 
${\Gamma_\mu}^{\dag}=\Gamma_\mu\,,\,{\Gamma_{2l+1}}^{\dag}=\Gamma_{2l+1}$.
Thus, we may write with all indices different 
$\mu_1\ne\nu_1\ne\ldots\ne\mu_{l-1}\ne\nu_{l-1}\ne\alpha\ne\beta$,
\be
\Gamma_{\mu_1}\Gamma_{\nu_1}\ldots\Gamma_{\mu_{l-1}}\Gamma_{\nu_{l-1}}
\propto
\epsilon_{\mu_1\nu_1\ldots\mu_{l-1}\nu_{l-1}\alpha\beta}\Gamma_{2l+1}
\Gamma_\alpha\Gamma_\beta
\quad.
\label{pfaffi}
\ee
Antisymmetrising the $(l-1)$ pairs of gammas this leads to
\be
\Sigma_{\mu_1\nu_1}\Sigma_{\mu_2\nu_2}\ldots\Sigma_{\mu_{l-1}\nu_{l-1}}
\propto
\epsilon_{\mu_1\nu_1\ldots\mu_{l-1}\nu_{l-1}\alpha\beta}\Gamma_{2l+1}
\Sigma_{\alpha\beta}
\quad,
\label{pfaffii}
\ee
an expression which is symmetric in the $(\mu\nu)$ pairs which are all 
different.
To check that the definition \eq{defmulti} for the $(2l-2)$ bracket is indeed 
$so(2l)$-valued, we notice that the $so(2l)$ commutators
$[\Sigma_{\mu\nu},\Sigma_{\rho\sigma}]\equiv 
iC^{\lambda\kappa}_{(\mu\nu)(\rho\sigma)}\Sigma_{\lambda\kappa}$
are non-zero only if
the pairs $(\mu\nu)\,,\,(\rho\sigma)$ have one (and only one) 
index in common.
Thus, the only non-zero $(2l-2)$-brackets have the form
$[\Sigma_{i_1k},\Sigma_{i_2k},\ldots,\Sigma_{i_{2l-2}k}]$ 
where all indices $i$ are different.
Since the ordinary product of such $\Sigma$'s sharing an index 
is already antisymmetric, we 
find that (cf. \eq{prooftheoremi})
\be
\eqalign{
[\Sigma_{i_1k},\Sigma_{i_2k},\ldots,\Sigma_{i_{2l-2}k}]&
\propto
C^{j_1j_2}_{(i_1k)(i_2k)}\ldots C^{j_{2l-3}j_{2l-2}}_{(i_{2l-3}k)(i_{2l-2}k)}
\{\Sigma_{j_1j_2},\ldots,\Sigma_{j_{2l-3}j_{2l-2}}\}
\cr
&
\propto
{\omega_{i_1k,\ldots,i_{2l-2}k}}^{\alpha\beta}_{\ \mdot}
\Gamma_{2l+1}\Sigma_{\alpha\beta}
\quad,\cr}
\label{reducible}
\ee
and we may now use the chiral projectors ${1\over 2}(1\pm\Gamma_{2l+1})$ to 
extract from the reducible $2^l\times 2^l$ representation $\Sigma_{\mu\nu}$ 
its two irreducible
$2^{l-1}\times 2^{l-1}$ components.


\section{Higher-order simple Lie algebras and their complete \break 
BRST operator}

The case of the exceptional algebras requires more care, and we shall not 
discuss here their realization.
We may nevertheless state the following

\medskip
\noindent
{\bf Theorem 5.1}\quad ({\it Classification theorem for higher-order simple 
algebras})

Given a simple Lie algebra $\g$ of rank $l$, there are $(l-1)$ 
$(2m_i-2)$-higher-order simple  
algebras associated with $\g$.
They are given by the $(l-1)$ Lie algebra cocycles of order $(2m_i-1)>3$
which may be obtained from the $(l-1)$ symmetric invariant polynomials 
on $\g$ of order $m_i>m_1=2$.
The $m_1=2$ case (Killing metric) 
reproduces the original simple Lie algebra $\g$; 
for the other $(l-1)$ cases, the skew-symmetric
$(2m_i-2)$-commutators define an element of $\g$ by means of the 
$(2m_i-1)$-cocycles. These higher-order structure constants 
(as the ordinary structure constants with all indices are
written down)
are fully antisymmetric and satisfy, by virtue of being Lie algebra cocycles,
the generalised Jacobi condition \eq{jaccond}.

\medskip
\noindent
{\it Remark.}\quad

It may be checked explicitly that the coordinate definition of the cocycles 
${\omega_{\ind i{2m-2}}}^\sigma_\mdot$ 
and the invariance condition \eq{invariance}\ for their associated invariant 
polynomials entail the GJI.
Indeed, the $l.h.s$ of \eq{jaccond} (for $s=2m-2$) is, using \eq{defcocycle}, 
equal to
\be
\eqalign{&
\epsilon^{j_1\ldots j_{4m-5}}_{i_1\ldots i_{4m-5}}
{\omega_{j_{1}\ldots j_{2m-2}\mdot}}^\rho
\epsilon^{l_1\ldots l_{2m-3}}_{j_{2m-1}\ldots j_{4m-5}}
C^{p_1}_{\rho l_1}\ldots C^{p_{m-1}}_{l_{2m-4}l_{2m-3}}
{k_{p_1\ldots p_{m-1}}}_\mdot^\sigma
\cr
=&
(2m-3)!
\epsilon^{j_1\ldots j_{2m-2} l_1 \ldots l_{2m-3}}_{i_1\; \ldots \;i_{4m-5}}
{\omega_{j_{1}\ldots j_{2m-2}\mdot}}^\rho
C^{p_1}_{\rho l_1}\ldots C^{p_{m-1}}_{l_{2m-4}l_{2m-3}}
{k_{p_1\ldots p_{m-1}}}_\mdot^\sigma
=0\quad,
\cr}
\label{nuevath}
\ee
which is zero since if 
${\omega_{\ind j{2m-2}}}^\rho_\mdot$ is a $(2m-1)$-cocycle [\eq{defcocycle}]
\be
\epsilon^{j_1\ldots j_{2m-1}}_{i_1\ldots i_{2m-1}}
C^{\nu}_{j_1 \rho}{\omega_{j_2\ldots j_{2m-1}}}^\rho_\mdot=0\quad,
\label{VIxii}
\ee
which follows from Lemma 3.1.

There is a simple way of expressing the above results making use of the 
Chevalley-Eilenberg \cite{CE} formulation of the Lie algebra cohomology.
For the standard case, we may introduce the BRST operator
\be
s=-{1\over 2}c^ic^j{C_{ij}}^k_\mdot{\partial\over\partial c^k}\quad,
\quad s^2=0\quad,
\label{BRST}
\ee
with $c^ic^j=-c^jc^i$
(in a graded algebra case, the $c$'s would have a grading opposite to that of 
the associated generators).
Then,
$sc^k=-{1\over 2}{C_{ij}}^k_\mdot c^ic^j$
(Maurer-Cartan eqs.) and the nilpotency of $s$ 
is equivalent to the JC \eq{jacid}.
In the present case, we may describe all the previous results by introducing 
the following generalisation:

\medskip
\noindent
{\bf Theorem 5.2}\quad ({\it Complete BRST operator for a simple Lie algebra})

Let $\g$ be a simple Lie algebra.
Then, there exists a nilpotent associated
operator given by the odd vector field
\be
\eqalign{
s=
&
-{1\over 2}c^{j_1}c^{j_2}{\omega_{j_1j_2}}^\sigma_\mdot
{\partial\over\partial c^\sigma}
-\ldots-
{1\over (2m_i-2)!}c^{j_1}\ldots c^{j_{2m_i-2}}
{\omega_{j_1\ldots j_{2m_i-2}}}^\sigma_\mdot
{\partial\over\partial c^\sigma}
-\ldots
\cr
&
-
{1\over (2m_l-2)!}c^{j_1}\ldots c^{j_{2m_l-2}}
{\omega_{j_1\ldots j_{2m_l-2}}}^\sigma_\mdot
{\partial\over\partial c^\sigma}
\equiv s_2+\ldots+s_{2m_i-2}+\ldots +s_{2m_l-2}
\;,
\cr}
\label{HOBRST}
\ee
where 
$i=1,\ldots,l\,,\,
{\omega_{j_1j_2}}^\sigma_\mdot\equiv{C_{j_1j_2}}^\sigma_\mdot$
and 
${\omega_{j_1\ldots j_{2m_i-2}}}^\sigma_\mdot$
are the corresponding $l$ 
($c$-number) higher-order cocycles.
The operator $s$ will be called the complete BRST operator associated with 
$\g$.

\medskip
\noindent
{\it Proof}:

The nilpotency of $s$ encompasses, in fact, the JC and
the $(l-1)$ GJC's which have to be satisfied, respectively, 
by the $\omega$'s which determine the standard BRST operator 
\eq{BRST} and the 
$(l-1)$ higher-order BRST operators; all the cohomological information on $\g$ 
is contained in the complete BRST operator.
The GJC's come from the squares of the individual terms
$s_{p}^2$, the crossed products $s_{p} s_{q}$ not contributing since the 
terms $s_{2m_i-2}$ are given by Lie algebra $(2m_i-1)$-cocycles.
To see this, we first 
notice that there are no $\omega$'s with an even 
number of indices ($s$ is an odd operator).
Consider now a mixed product $s_p s_q$ ($p$ and $q$ even).
This is given by
\be
\eqalign{
s_ps_q
&\propto
{\omega_{i_1\ldots i_{p}}}^\rho_\mdot 
c^{i_1}\ldots c^{i_{p}}
{\omega_{j_1\ldots j_{q}}}^\sigma_\mdot
[\sum_{l=1}^q (-1)^{l+1}\delta_\rho^{j_l}c^{j_1}\ldots \hat c^{j_l}\ldots 
c^{j_{q}}]{\partial\over\partial c^\sigma}
\cr
&=
q{\omega_{i_1\ldots i_{p}}}^\rho_\mdot
{\omega_{j_1\ldots j_{q}}}^\sigma_\mdot \delta_\rho^{j_1}
c^{i_1}\ldots c^{i_{p}}
c^{j_2}\ldots c^{j_q}{\partial\over\partial c^\sigma}
\cr
&=
q{\omega_{i_1\ldots i_{p}}}^\rho_\mdot
{\omega_{\rho j_2\ldots j_{q}}}^\sigma_\mdot
c^{i_1}\ldots c^{i_{p}}
c^{j_2}\ldots c^{j_q}{\partial\over\partial c^\sigma}\quad,
\cr}
\label{mixed}
\ee
where the term in 
${\partial\over\partial c^\rho}{\partial\over\partial c^\sigma}$
has been omitted since ($p$ and $q$ being even)
it cancels with the one coming from $s_q s_p$.
Recalling now expression \eq{defcocycle} it is found that \eq{mixed} is zero 
because of \eq{VIxii}, which in the present language reads $s_{p}s_2=0$.
Thus, 
$s^2 =s^2_2+\ldots+ s^2_{2m_i-2}+\ldots+s^2_{2m_l-2}=0$, each of the $l$ terms 
being zero separately as a 
result of the GJC \eq{jaccond}, {\it q.e.d.}


\section{Concluding remarks}
Many questions arise now that require further study. From a physical point 
of view it would be interesting to find applications of 
these higher-order Lie algebras to know whether the cohomological restrictions 
which determine and condition their existence have a physical significance.
Lie algebra cohomology arguments have already been very useful 
in various physical problems 
as \eg, in the description of anomalies \cite{TJZW} or in the 
construction of the Wess-Zumino terms required in the action of extended 
supersymmetric objects \cite{AZTO}.
In the form \eq{HOBRST}, the above formulation of the higher algebras 
has a resemblance with the closed 
string BRST cohomology and the SH algebras \cite{LAST,LAMA} 
relevant in the theory of graded string field products \cite{WZ,ZIE}
(see also \cite{LZ}).
Note, however, that because of the cocycle form of the $\omega$'s, the
GJI's are not modified as already mentioned in the introduction.
In the SH algebras such a modification is the result of having, for instance, 
terms lower than quadratic in \eq{HOBRST} 
(with the appropriate change in ghost grading).

Due to their underlying BRST symmetry, similar structures appear in the 
determination of the different gauge structure tensors through the 
antibrackets and the master equation in the Batalin-Vilkovisky formalism (for 
a review, see \cite{MH,GPS}), where violations of the JI are 
also present (the Batalin-Vilkovisky antibracket is a two-bracket, but 
higher-order ones may also be considered \cite{BDA}).

Other questions may be posed from a purely mathematical point of view.
As the discussion in sec. 4 shows, a representation of a
simple Lie algebra may not be a representation for the associated 
higher-order Lie algebras.
Thus, the representation theory of higher-order algebras requires a separate 
analysis.
Other problems may be more interesting from a structural point of view as, 
for instance, the contraction theory of higher-order Lie algebras (which will 
take us outside the domain of the simple ones), as 
well as the study of the non-simple higher-order algebras themselves and 
their cohomology.
These, and the generalisation of these ideas to superalgebras (for which there 
exist simple finite dimensional ones with zero Killing form)
are problems for further research.

\section*{Acknowledgements}
The authors wish to thank J. Stasheff for helpful correspondence and his 
comments on the manuscript and T. Lada 
for a copy of \cite{LAMA}.
This research has been partially sponsored by the Spanish CICYT and DGICYT 
(AEN 96-1669, PR 95-439).
Both authors wish to thank the kind hospitality extended to them at DAMTP.
The support of St. John's College (J.A.) and an FPI grant from the Spanish 
Ministry of Education and Science and the CSIC (J.C.P.B.) are also gratefully 
acknowledged.


\end{document}